# Research on rolling friction's dependence on ball bearings' radius

**GRZEGORZ M. KOCZAN**
**JULIUSZ ZIOMEK**

**ABSTRACT:** There are two alternative historical laws of rolling resistance formulated by French scientist *Coulomb* and *Dupuit*. It has been decided to verify experimentally again, which of these laws describes freely rolling ball bearings on a hard surface better. An inducement to carrying out the measurements was the idea of the constant thickness of roadbed, which is consistent with *Dupuit's* theory. Measurements have been done using the damped oscillations in the pendulum bearings. Results have shown better consistency with the *Coulomb's* theory with small, but measurable deviations. These deviations were successfully explained by the so called "Cobblestones model". Parameters designated by this model have been successfully verified by the surface roughness's profile measurement. An additional theoretical aspect of this work is distinguishing two types of rolling friction force: dynamical and kinematical in an analogy to two types of specific heat capacity in the thermodynamics of gases.

The invention of the wheel in 4th millennium BC in Mesopotamia enabled humanity to decrease the friction forces by two orders of magnitude, but have not eliminated them completely. Even during the movement of the wheel without slipping, there is a phenomenon called rolling friction. It is in general present on the periphery of the wheel, where it touches the ground, but also in roller bearings. The importance of the roller bearings is indisputable [1]. However, we may not be aware, how catastrophic a malfunction of this element can be. A tragic example of that is the greatest, in terms of the number of casualties (183 killed), flight catastrophe in Polish aviation, which happened in 1987 in The Kabacki Forest. The cause of the crash of plane Ił-62M "Tadeusz Kościuszko" was faulty of roller bearings of the engine's shaft. [1]

The pioneer of the research of rolling resistance was a French scientist Charles-Augustin de Coulomb. Beside his most famous law concerning electrostatics, he had also formulated a law of rolling friction. According to his studies published in year 1785, during which his he analysed the movement of wooden cylinders (guaiac and elm ones) on the oak planks, the force of rolling friction is proportional to the magnitude of contact force $N$ and inversely proportional to the radius $R$ of rolling solid [2]. It is expressed by Coulomb's formula for the force of rolling friction $T_f$ [3] :

$$T_f = \frac{f}{R} N \tag{1}$$

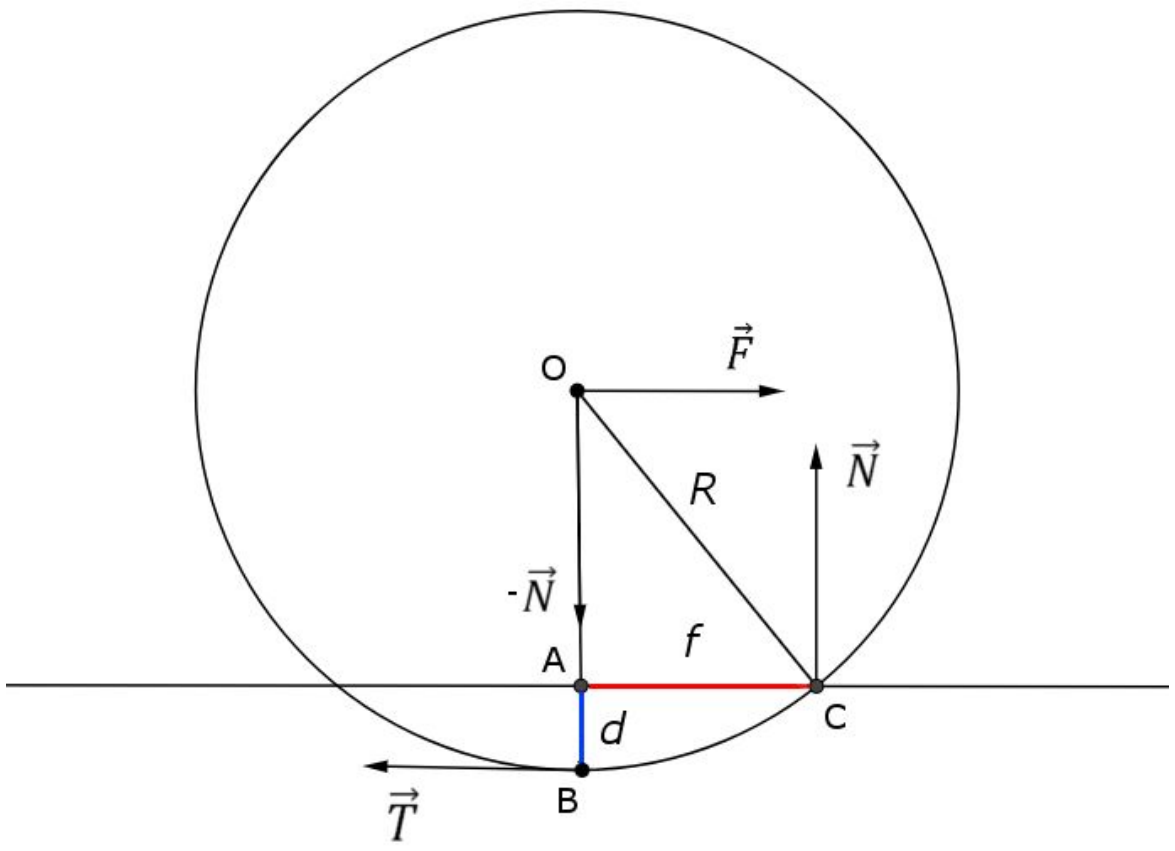

Figure 1. A schematic figure of basic forces acting on a rolling solid

where: $f$ – special coefficient of rolling friction with the unit of length. During movement, coefficient $f$ can be interpreted as a lever arm of the ground reaction force [4]:

[1] Earlier in 1980, plane Ił-62 "Mikołaj Kopernik" with 87 people on board crashed, whereas on 11 March 1990, thanks to the sensors of the engine's vibration, a catastrophe of Ił-62M was avoided.

$$M_f = f \cdot N \tag{2}$$

Besides an apparent mathematical equivalence, equation (2) is a bit more general than equation (1). Analogically, in the energetic considerations, it is easiest to consider the work done by the torque of rolling friction in the rotational movement:

$$W_f = -M_f \cdot \theta = -f \cdot N \cdot \frac{s}{R} = -T_f \cdot s \tag{3}$$

where: $\theta$ – angle of solid's rotation, $s$ – distance travelled.

This approach will let us avoid apparent paradoxes of zeroing of work done by rolling friction (because of the lack of slipping or the work in progressive and rolling movement cancelling each other out). Besides the crucial role that coefficient $f$ has, academics often define via an analogy with sliding friction, a unitless coefficient of friction $\mu_f$, as [5]:

$$\mu_f = \frac{f}{R} \qquad \mu_f = \frac{T_f}{N} \tag{4}$$

The first formula can be treated as an equivalent form of Coulomb's law (1). It turned out that in 1837 another French scholar proposed alternative law [6]. Engineer and economist Arsène Jules Dupuit, by examination of steel-shod wooden wheels, found an empiric relation between friction and inverse of a square root of its radius [7, 8]:

$$\mu_f = \sqrt{\frac{D}{R}} \quad \Leftrightarrow \quad f = \sqrt{DR} \tag{5}$$

where $D$ is a unit-length constant depending on used materials.

Coulomb and Dupuit formulas are not the only ones that are being analysed by modern scholars in terms of empiric and theoretical relations. As an example, Palmgren's formula is describing drag torques in roller bearings [9]. Friction during rolling is a broad issue, requiring miscellaneous material distinctions to cases of soft or firm, elastic or plastic, rough or covered with grease and so on[2]. Taking this into account, the range of works will be limited to the most famous approach of Coulomb and the empiric result of Dupuit, for which it is possible to give simple geometrical interpretations (with possibilities of small modifications).

The theory of Coulomb assumes that for given materials, the $f$ *is* constant regardless of the radius of rolling solid. Omitting the simplified model of uniform unevenness of surface or body, it is hard to visualize a different explanation of Coulomb's theory. An alternative geometrical magnitude that can be naturally constant is the thickness of surface layer $D$ or in other words – the depth of penetration of the body into the surface (depending on the roughness or elasticity). It is the idea of constant $D$ that is an inducement to the measurement of rolling friction of unloaded bearing balls in this work. This idea predicts the value of Dupuit constant $D$. It results from the Pythagoras theorem, applied to the triangle AOC (Fig. 1) for small $d \ll R$:

$$f = \sqrt{R^2 - (R-d)^2} = \sqrt{2Rd - d^2} \approx \sqrt{2Rd} \quad \Rightarrow \quad D \approx 2d \tag{6}$$

The practical aim of this work is to settle between Coulomb's law of constant $f$ and Dupit law of constant $d$ in case of freely rolling bearing balls on the surface of painted metal. Because of the lack of known model that would clearly be consistent with one of the laws, this settlement is not that obvious. Up to date experimental research is ambiguous and often shows different exponential relationships [5, 6, 8].

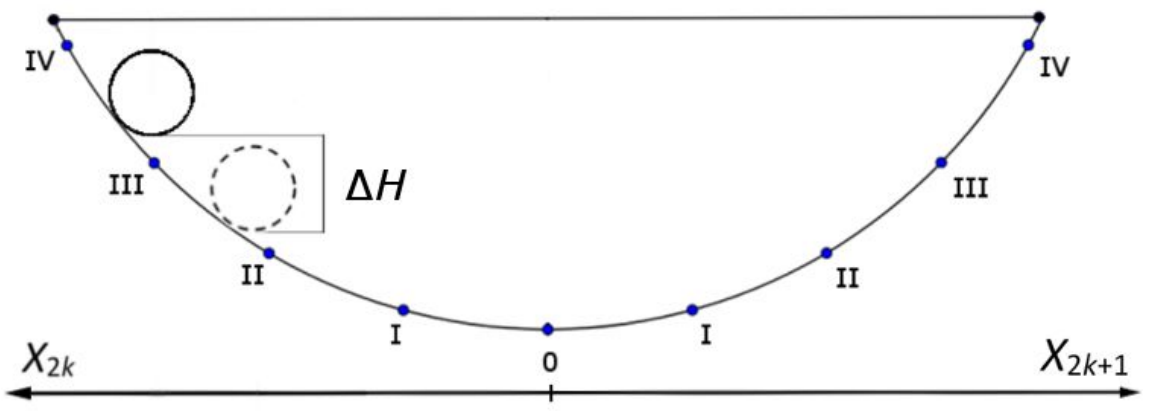

Figure 2. Scheme of experimental design, circular bowl that serves as a track for consecutive bearing balls

[2] The shape of the rolling body (spherical or cylindrical) might also be crucial.

## Methods of measurement and calculations

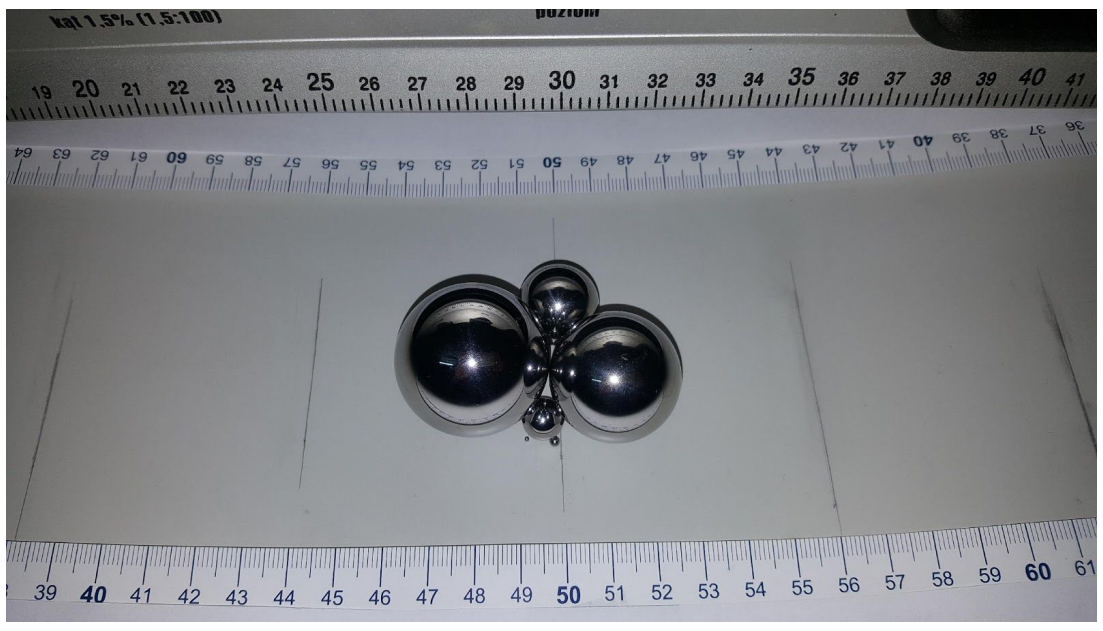

Figure 3. Photo of balls with diameters: 1.2, 2.5, 10, 21, 30, 34.9 mm inside a satellite dish

The experimental design consisted of 48 bearing balls with diameters in the range 1.2–34.9 mm (two in each size) and the satellite dish of elliptical shape, which was a fragment of a circle in its cross-section, which was proved by the examination of the profile of the dish. The position of the dish was set using a spirit level and stabilised so that rolling balls were not producing any vibrations. Two linear scales were glued to the dish (Fig. 3). Next, bearing balls were freely rolling down the dish one by one, each time reaching lower and lower heights, which happened due to the force of rolling friction performing work. Their movement was recorded with a camera and analysed in slow-motion on the recording so that one can clearly read the maximum distance from the centre that the balls reached after consecutive full swings. These distances (length measured on the arc of the dish) were marked as: $s_0$, $s_2$, $s_4$, …, $s_{2n}$ (distances of swings of odd numbers at the other side of the dish were not measured to avoid the error of parallax). Based on the real distances their horizontal projections were calculated (positive coordinates) $x_0$, $x_2$, $x_4$, …, $x_{2n}$. In the case of the used dish (sagitta of 38 mm on the distance of 240 mm), both lengths were approximately equal with a certainty of 1% ($x \approx s$). However, measuring every distance was not an effective approach for large balls, as they were performing a lot of swings before coming to a stop, and the changes between consecutive swings were insignificant. In this case a different method was used. On the dish, precisely 9 lines were drawn with a spacing of 60 millimetres. Line number 0 was exactly in the middle and four lines (I–IV) on the left and four (I–IV) on the right (Fig. 2). Next, during observations the number of full swings were calculated, until, at the point of maximum displacement from the centre, the ball's radius was exactly covering each consecutive line drawn on the dish. Numbers of full swings performed before crossing each line were marked as $n_i$, For example, for the second line – $n_{II}$. To avoid an error, balls which were close to the point where the experimental method was changed were measured using two methods (the ones with a diameter of 13 and 14 mm). Because the profile of the dish was known, so was the exact height $H_i$ of each of the lines, then by measuring the distance from the centre of the dish to the ball, it was possible to determine its height relative to the bottom of the dish and knowing the difference in height – losses of energy. And on the very law of conservation of energy are the calculations based, comparing the work done by rolling friction and the change in potential energy in the ball's highest position. Work of the rolling friction done from the equilibrium point to maximal displacement is given by the integral:

$$W_f = -\int \frac{f}{R} N \,\mathrm{d}s = -\int_0^{\varphi} \frac{f}{R} m g \cos\varphi \; r \,\mathrm{d}\varphi \tag{7}$$

where: $r = 732$ mm – radius of the dish's curvature, $\varphi$ – angular coordinate of the position of the ball in the dish.
The result of this integral is expressed as following:

$$W_f = -\frac{f}{R} m g x \approx -\frac{f}{R} m g s \tag{8}$$

where: $x$ – horizontal track of the ball. Thus, in approximation, the work of the rolling resistance is expressed by the same formula as for a horizontal surface. Multiple swings of the ball require, however, summing up either horizontal or real parts of the balls' displacements:

$$S = x_0 + 2x_1 + 2x_2 + \ldots + 2x_{2n-1} + x_{2n} \tag{9}$$

Numerical calculations have shown that equally precise summation can be done by the arithmetic and geometric sequences. An intermediate version will be shown – an approximation of geometric sequence by an integral:

$$S \approx 2\int_0^{2n} x_0 \left(\frac{x_{2n}}{x_0}\right)^{\frac{k}{2n}} \mathrm{d}k = \frac{x_0 - x_{2n}}{\ln(x_0 / x_{2n})} 4n \tag{10}$$

For comparison, a formula referring to the assumption of the arithmetic sequence is given as:

$$S = (x_0 + x_{2n}) \cdot 2n \tag{11}$$

By using sum $S$, the law of conservation of energy can be written in the following form:

$$\frac{f}{R} mg \cdot S = mg \cdot \Delta H \tag{12}$$

where $\Delta H$ is a decrement of height at the highest position of the ball. It leads to a complete formula for the coefficient $f$:

$$f = \frac{R \cdot \Delta H}{S} \tag{13}$$

Additionally, measurements of the roughness profile of the dish and balls were made. Based on the measurements, an average profile height $Rz$ and an average width of the profile's elements $RSm$ was noted, in accordance with PN-EN ISO 4287:1999 [10] standard. These values were measured to verify the further proposed "Cobblestone model".

## Results of measurements

For each of the balls, rolling friction coefficient $f$ were derived, using the described procedure. Coefficients corresponding to the same pairs of balls were averaged. Based on this, a depth of penetration $d$ was calculated and a unitless coefficient of rolling friction $\mu_f$. Averaged results for all the balls and the greater balls are shown in Table I.

**TABLE I. Results of measurements of rolling friction's parameters**

| | | |
|---|---|---|
| Radius range $R$ [mm] | 0.6 – 17.5 | 9 – 17.5 |
| Rolling friction coefficient $f$ [µm] | 15.2 ± 2.6 | 16.3 ± 1.0 |
| Depth of penetration $d$ [µm] | 0.009 – 0.092 | 0.0105 ± 0.0012 |
| Unitless coefficient of rolling friction $\mu_f$ | 0.001 – 0.017 | 0.00130 ± 0.00022 |

These results show the superiority of Coulomb's theory over Dupuit's in the measured range. Besides this, if one looks at the diagram on Fig. 5, the depth of penetration $d$, starting from range 9 mm seems as if it is constant. Values in Table I show, however, that relative standard deviation of $d$ is greater than $f$, even for large balls.

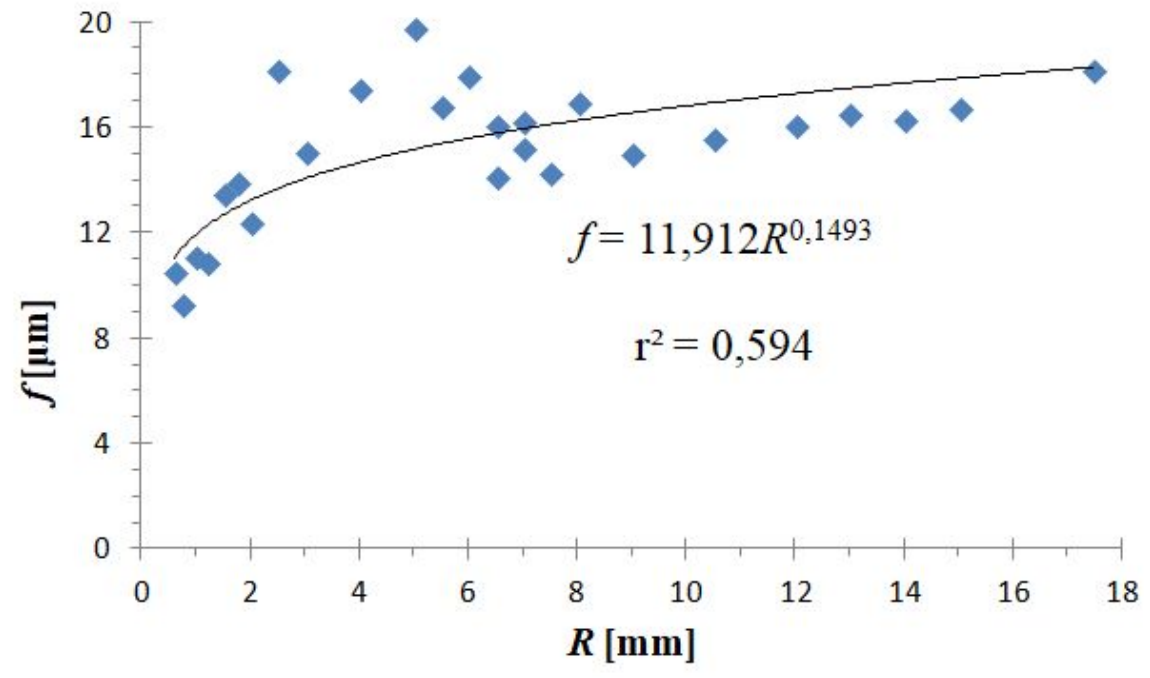

Figure 4. Relationship between $f(R)$, rolling friction coefficient and the radius of the ball

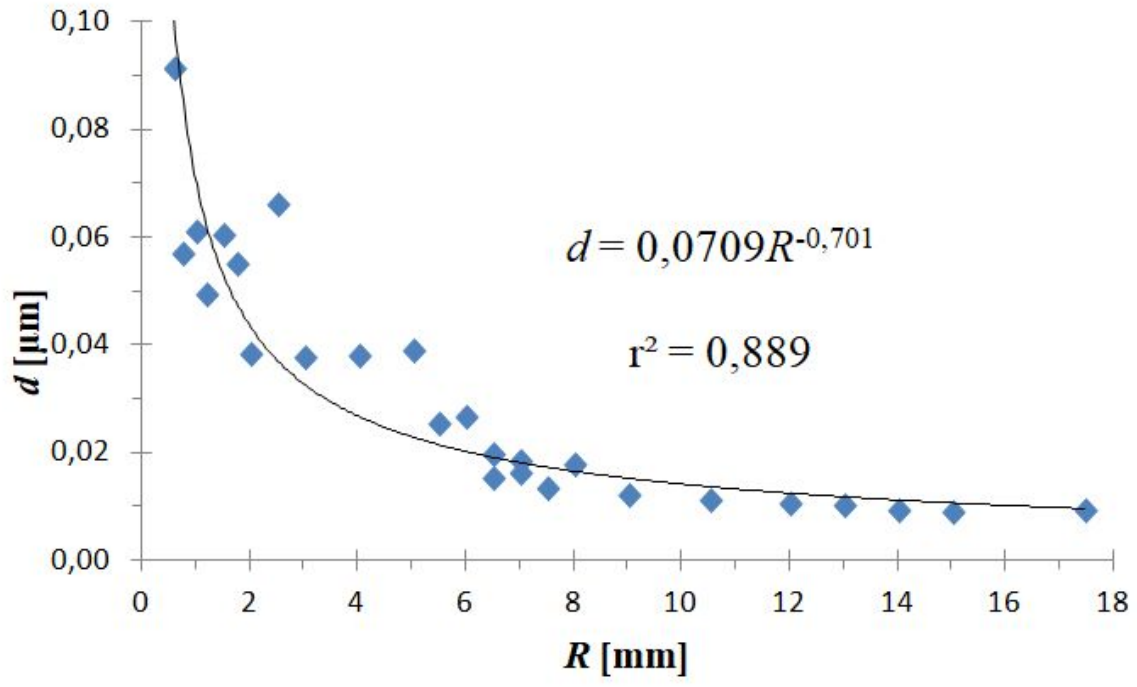

Figure 5. Relationship between $d(R)$, depth of penetration and the radius of the ball

Especially peculiar are also four measurement points of medium-sized balls (radius from 2 to 5 mm), which shown on the diagram in Fig. 5, form a next "shelf" of almost constant $d$. "Shelves" for great and medium balls during the conducting and analysis of the experiment, made it look as if the observations were more consistent with one of the theories, and then with the other. It was shifting a few times before it eventually settled on the Coulomb theory. However, the exponents of the trend lines on Figures 4, 5 and 6 deviate slightly from the desired values equal respectively 0, -1 and -1. These small deviations from the Coulomb theory can be explained by the "Cobblestone model".

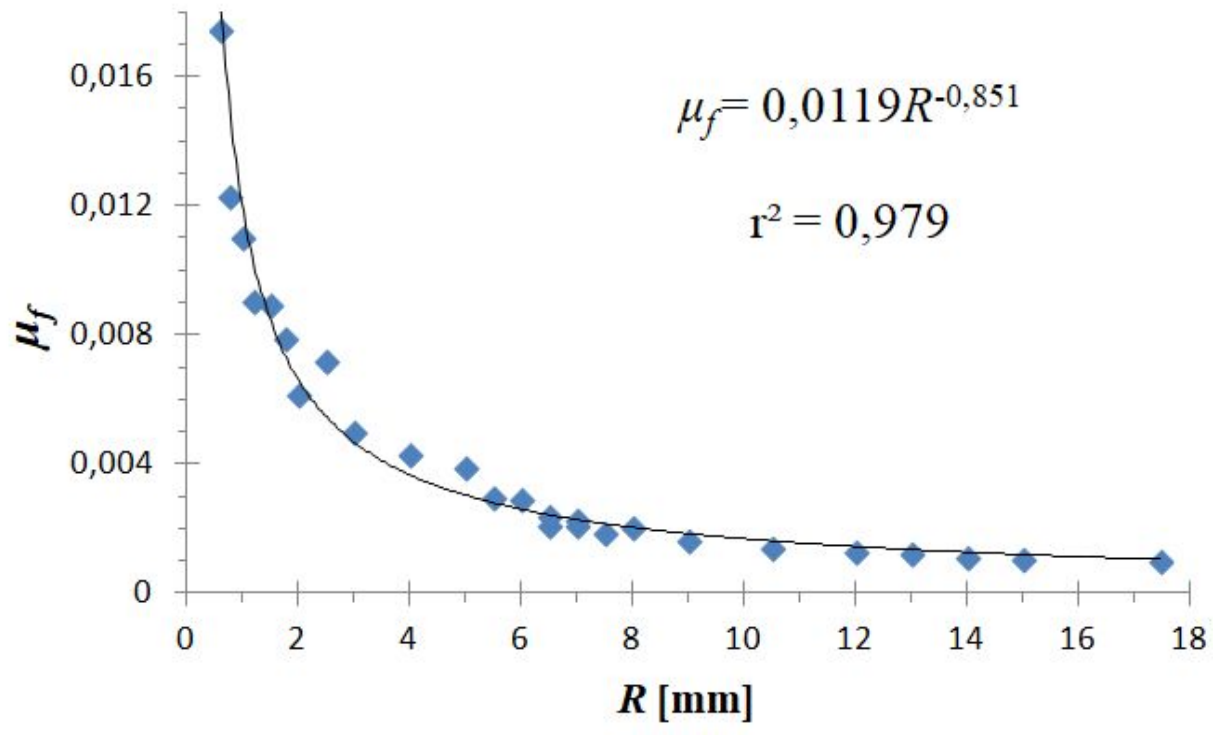

Figure 6. Relationship between $\mu_f(R)$ dimensionless friction coefficient and the radius of the ball

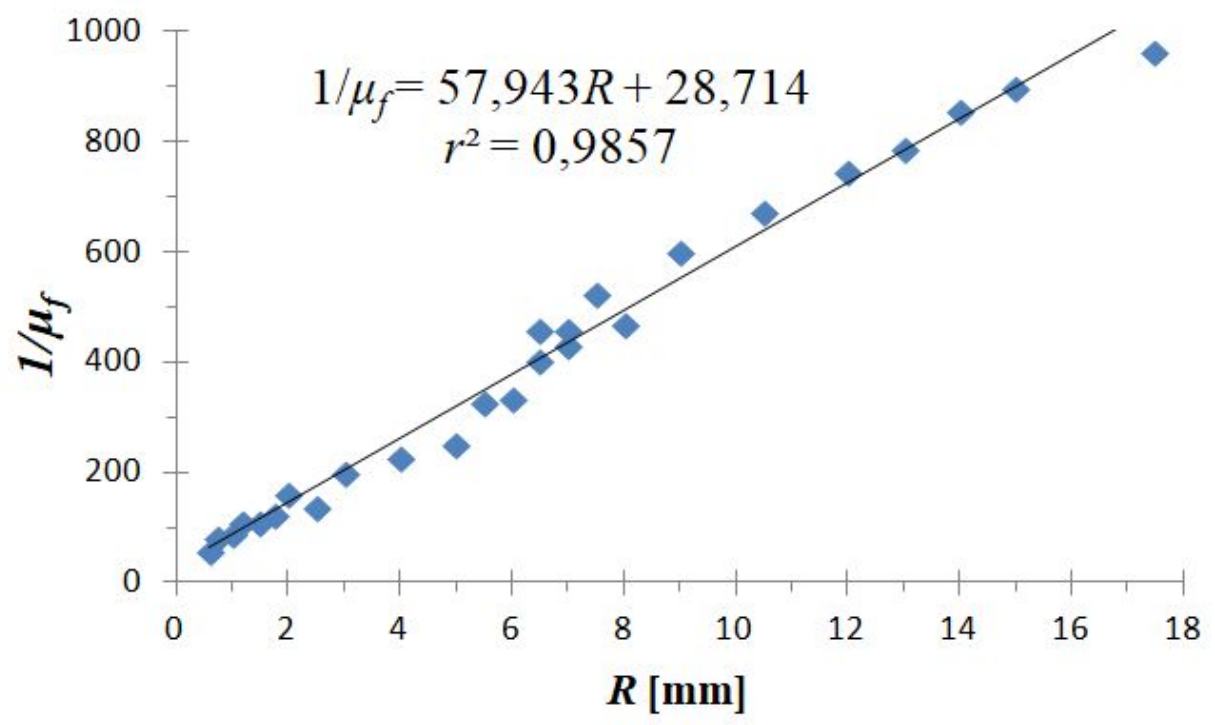

Figure 7. Relationship between $1/\mu_f(R)$, inverse of unitless friction coefficient and the radius of the ball

## The "Cobblestone model"

The law of Coulomb for the rolling friction predicts undisturbed inversely proportional relation $\mu_f = f/R$. However, after a horizontal translation to the left by some value $\rho$ this law can be expressed as $\mu_f = f/(R+\rho)$. Exponential regression of such relationships will give effectively instead of -1, a slightly higher exponent value, for example -0.85 as in the diagram on Fig. 6. To determine a new, additional parameter, one just needs to use linear regression for the relationship $1/\mu_f$ in respect to $R$ shown on the diagram on Fig. 7. The highest coefficient of determination $r^2 = 0.9857$ of the ones considered so far, indicates that the investigation is heading in the correct direction.

It turns out that this mysterious parameter $\rho$ can be interpreted as a radius of a "cobblestone" in the surface of the "Cobblestone model" (Fig. 8). Second parameter of this model is the distance $\lambda = |S_1S_2|$ between centres of curvatures' radiuses of neighbouring "stones". Based on the Pythagoras Theorem in the triangle $KLS_1$ height $h$, at which "stones" stick out of the surface can be calculated:

$$h = \rho - \sqrt{\rho^2 - (\lambda/2)^2} \approx \frac{\lambda^2}{8\rho} \tag{14}$$

where the approximation formula results from the transformation of the approximation formula of type (6) and is correct for $\lambda / 2 << \rho$. The most important difference, however, is the difference in height which occurs when the ball is rolling on the cobblestones. It can be calculated based on the triangle $KOS_1$:

$$\Delta h = R + \rho - \sqrt{(R+\rho)^2 - \frac{\lambda^2}{4}} \approx \frac{\lambda^2}{8(R+\rho)} \tag{15}$$

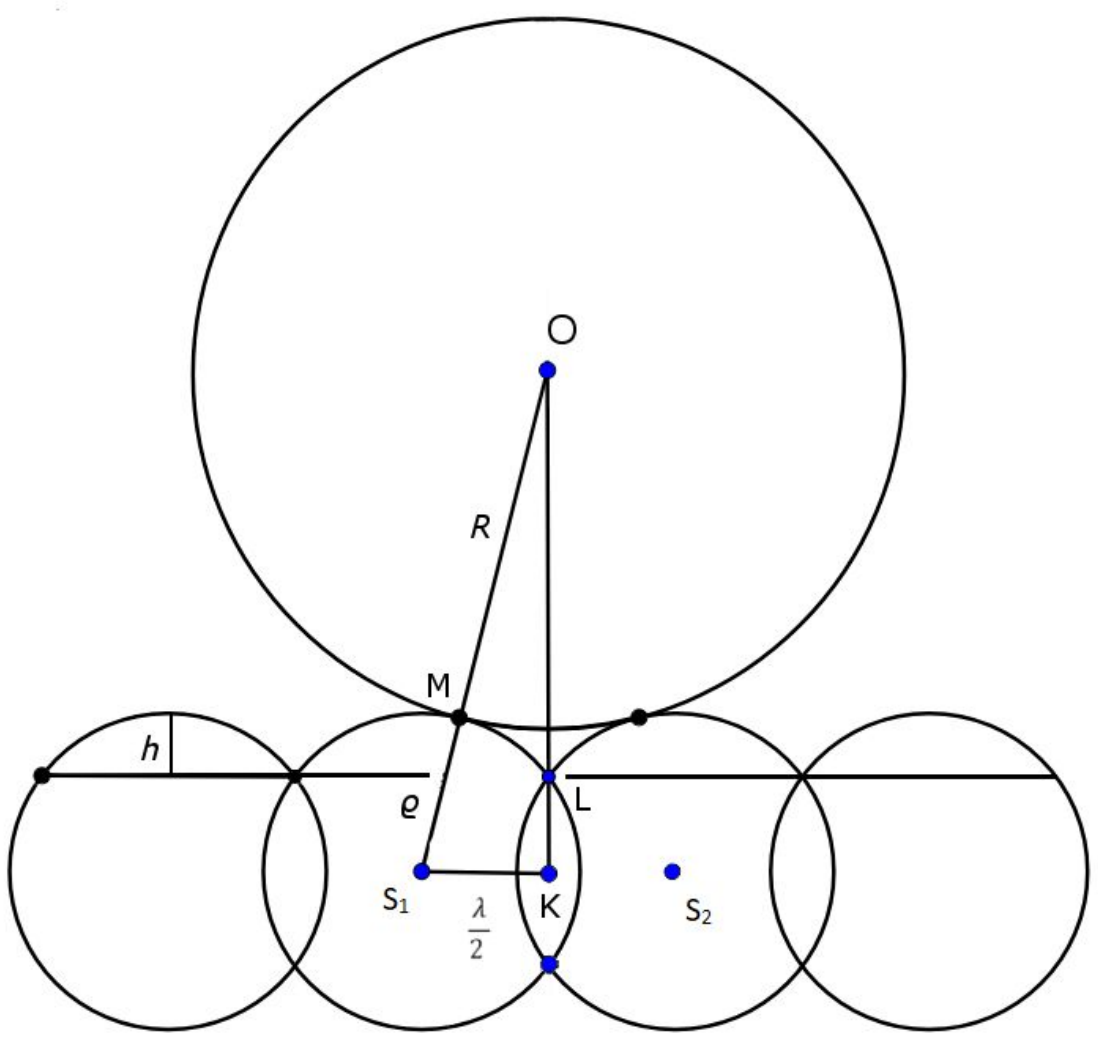

Figure 8. Schematic of the geometry of the "Cobblestone model"

This change in height during bouncing on the cobblestones relates to the loss of potential energy (or the work of the pressure force), which can be identified as the work of rolling resistance forces:

$$\frac{f}{R} N \cdot s = N \cdot \Delta h \cdot \frac{s}{\lambda} \tag{16}$$

where $s/\lambda$ is the number of bounces on the way $s$. The transformation of the two last formulas leads to a slightly modified Coulomb's law in the "Cobblestone model":

$$\mu_f = \frac{\lambda / 8}{R + \rho} = \frac{f}{R} \tag{17}$$

In the considered model, the coefficient $f$ is only asymptotically constant for large balls ($f \approx \lambda/8$), whereas for the small ones it has a slightly smaller value (as on diagram in Fig. 4). Parameters of the "Cobblestone model" can be also compared with the parameters of Dupuis's law. Based on the formulas (6), (15), (16), (17) the depth $d$ constitutes for around 1/16 of the difference in height $\Delta h$ in bouncing of the ball, which is dependent on $R$ and smaller than the height of the profile $h$. These approximate relations of parameters can be summarized as:

$$f \approx \lambda/8 \quad , \quad d \approx \Delta h/16 \quad , \quad d << h \tag{18}$$

However, the radius of a "cobblestone" $\rho$ as a parameter of the model with the opposite sign can be interpreted as an intersection of horizontal axis on the linear diagram of $1/\mu_f$ against $R$ (diagram on Fig. 7). Therefore, it is clearly visible that it has a small value, however, it is also bigger than any other parameter of friction concerned in this paper.

Linear regression of the experimental data on the diagram on Fig. 7 yields parameters of the model given in Table II.

**TABLE II. Comparison of experimental parameters of the "Cobblestone model" and the results of the measurements of the profile of the dish's surface roughness**

| Unit [μm] | Cobblestone model | Roughness measurement |
|---|---|---|
| Unevenness radius (of stone) | $\rho$ = 496 ± 197 | undefined |
| Unevenness width | λ = 138.1 ± 3.4 | $RSm$=301 ± 122 |
| Height of roughness of the surface | $h$ = 4.8 ± 1.9 | $Rz$=6.44 ± 0.53 |

The derived parameters were compared with the results of the measurements of the roughness' profile for the dish, calculated based on eight different measurements.

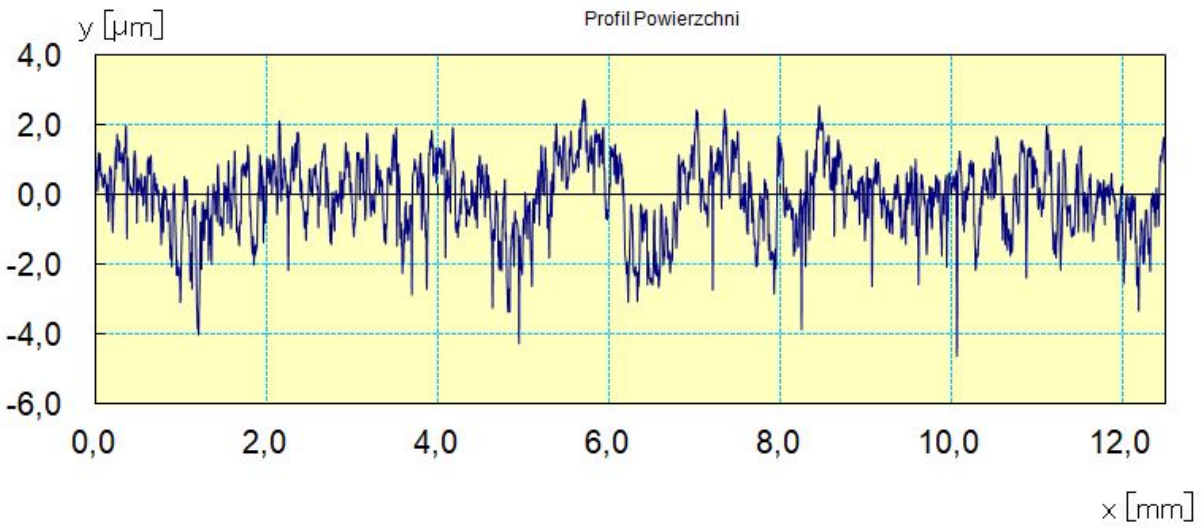

Figure 9. The profile of the roughness of the dish's surface

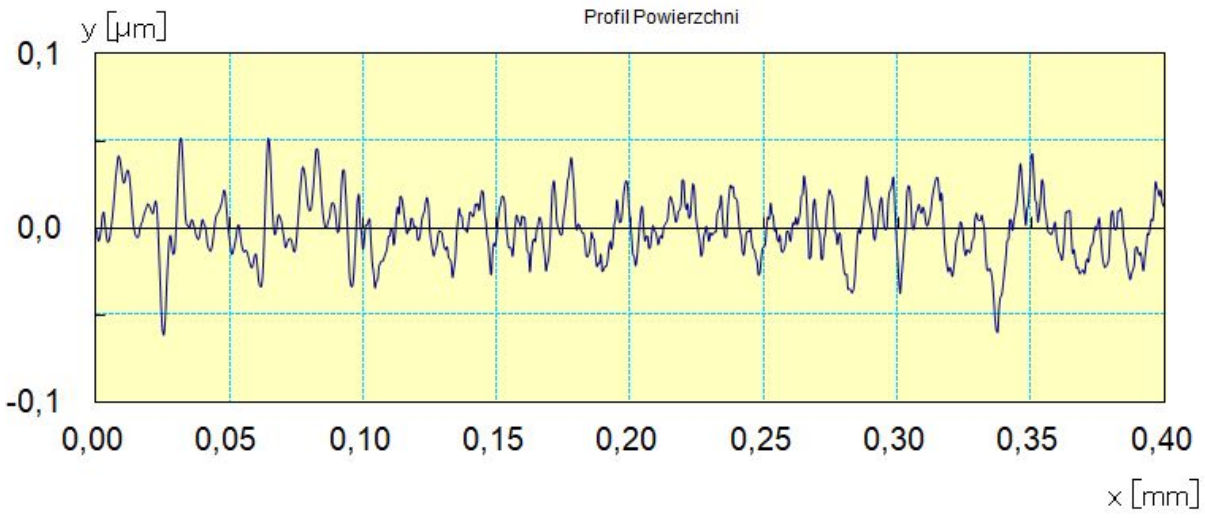

Figure 10. The profile of the roughness of the surface of a bearing ball

Also, six measurements of roughness of two balls were conducted, which parameters were almost an order of magnitude smaller than these ones of the dish ($RSm$ = 16.9 ± 3.2 µm, $Rz$ = 0.112 ± 0.022 µm). Obtained divergence $RSm$ of the balls with the rolling friction coefficient $f$, besides lapidary interpretation is probably random. If the cross-section of the ball was a polygon, then the coefficient of friction would be four times smaller than its side (and not equal to it). The factor of 1/4 in the model of a polygon for the ball (or in the "sharp teeth" model of the surface) and the factor of 1/8 in the "Cobblestone model" results from its kinetic characteristic in contrast to a quasi-static situation on Fig.1.

Based on that, it can be stated that the measurement of the roughness verifies the quantitative predictions of the "Cobblestone model" for the surface of the dish ($h \approx Rz$, $\lambda \sim RSm$) with negligible roughness of the balls.

**The force of rolling friction in a kinematical and dynamical approach**

In considerations about rolling friction, a formula for the angular force plays the main role. However, it does not give directly the information about the force of the rolling friction, which is acting on the body. It turns out that this force can be derived in two complementary, but not equivalent ways. Assuming designations as in Fig.1: $f$ – force putting solid into motion, $T$ – a total force of friction, $N$ – total pressure of the solid (considering weight and load), we end up with the following equations of motion:

$$ma = F - T \tag{19}$$

$$I\varepsilon = TR - Nf \tag{20}$$

Additionally, it is assumed that no slipping occurs[3]:

$$a = \varepsilon R \tag{21}$$

From equations (20) i (21), force of friction as a linear function of acceleration can be derived:

$$T = \frac{I}{R^2}a + \frac{f}{R}N = T_a + T_{f|a} \tag{22}$$

[3]Implicitly we assume thereby that static friction dominates over the rolling friction $\mu_s \gtrsim \mu_f$.

and the absolute terms of this expression can be treated as the kinematic force of rolling friction $T_{f|a}$. Whereas, if equation (19) is to be considered, we will receive a linear relationship between friction and the applied force:

$$T = \frac{I}{I + mR^2} F + \frac{mR^2}{I + mR^2} \frac{f}{R} N = T_F + T_{f|F} \tag{23}$$

with the absolute term equal to the dynamic force of rolling friction $T_{f|F}$. The occurrence of two types of rolling friction's force is surprising, but it has its quantitative and qualitative interpretation. Kinetic force of rolling friction acts when the outside force $F$ keeps the constant movement of the solid ($v$ = const, $a$ = const = 0). Whereas, dynamic force, smaller in value, acts in free braking, only by the resistance forces ($F$ = const = 0, $a$<0). Mathematically both types of friction forces can be defined as follows:

$$T_{f|a} = f\left(\frac{\partial T}{\partial f}\right)_a = \frac{f}{R} N \tag{24}$$

$$T_{f|F} = f\left(\frac{\partial T}{\partial f}\right)_F = \frac{1}{\alpha + 1} \frac{f}{R} N \tag{25}$$

where the partial derivatives are computed with a constant value of the variable in the lower index, whereas $\alpha$ is the coefficient of the moment of inertia $I = \alpha m R^2$. Formulas (24) and (25) are fully analogical to the definition of the specific heat capacity of gas with constant pressure and specific heat capacity with constant volume defined based on the specific entropy. It needs to be stressed that all the friction forces considered here (22)–(25) are of a static type and does not perform work (negative work in the progressive movement is balanced by the positive work in rolling motion). However, formally the value of the work of the force in progressive motion (23) is equal to the real work of the rolling friction angular force. This fact, in some way, marks out the kinetic force of the rolling friction.

In this work, it was assumed that all forces are applied at point B (Fig. 1). Any movement of these forces or part of them to point C would only complicate most of the formulas, without any practical gain for the interpretation of the theory. It is true that this manipulation would give the possibility of explaining the rolling friction as real sliding friction in point C, however, it would create further problems. Most of all, it would yield an inequivalent theory of rolling friction in the sense of different acceleration of the body.

## Conclusion

The measurements of rolling of unloaded bearing balls on a solid surface showed a better correlation with known Coulomb's law than with the Dupuit's Law. Regardless of this, a growing trend of $f$ in respect to $R$ was observed – inconsistent with the first law, but much smaller than what the latter predicts. The observed deviation was explained by the "Cobblestone model" (compare Fig. 8 with Fig. 11), which in some sense parametrise the roughness of the surface. The derived parameterization turned out to be consistent with the measurement of roughness's profile. Therefore, "Cobblestone model" has a double status. On one hand, it is a simple mechanical model of roughness's envelope for Coulomb's law. On the other hand, formula (17) can be treated as a sort of generalised Coulomb's law, allowing for better analysis of the experimental data. It applies to the linear regression for the inverse of the unitless coefficient of rolling friction $1/\mu_f$ with respect to the radius $R$. Conducted measurements have shown a dominating role of surface's roughness for the rolling resistance of solid and unloaded bodies. Of course, this result will not be adequate in case of huge load, but it does shed some light on the crucial impact of roughness on the rolling friction in reference to, for example, the elastic hysteresis. Theories based on the elastic hysteresis in the rolling friction are very complicated and as for now, have not yielded any simple results, clearly consistent with the experimental data [6].

In-depth analysis of the ways of separating the rolling friction force out of total friction has led to the identification of two such forces: kinetic (24) and dynamic (25). The expression for the dynamic force (25), acting for example, during slowing down of an accelerated bike, can also be treated as some modification of Coulomb's law. Free of such modifications is the approach using only the angular force of rolling friction (2). However, problems, in which rolling friction is directly considered, without mentioned clarification, might be difficult to solve. Authors of this paper claim that such ambiguity was present in problem 1.4 on the Advanced Physics Polish New Matura 2008 exam (Polish high school final exam). It is this problem that was an incentive to formulate definitions (24) and (25).

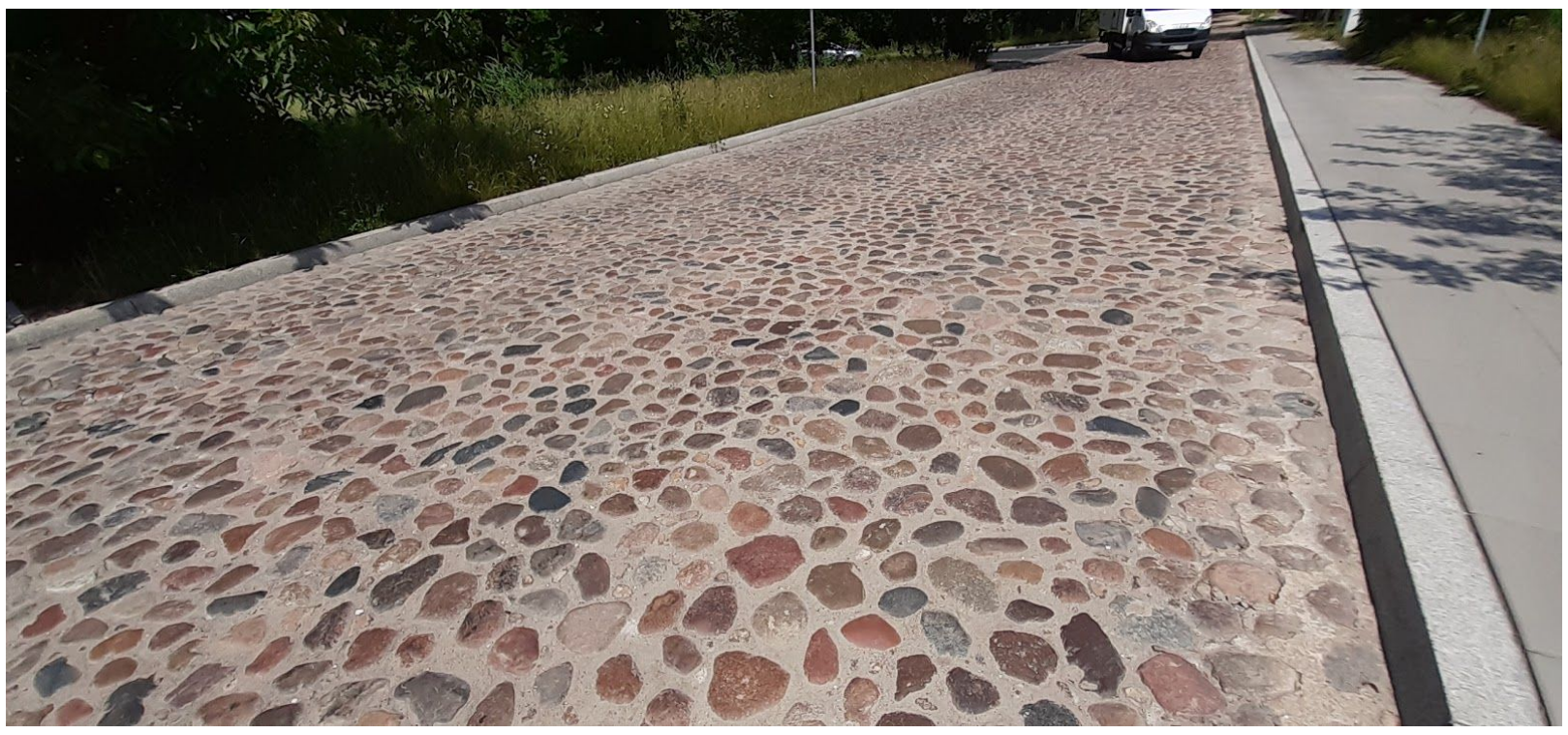

Figure 11. Cobblestone on Wolicka Street in Warsaw (Poland). In some countries, this type of cobblestone is literally called "cats heads". In Slovenia, they say "Mačje glave", in Poland "kocie łby". Therefore, the original name of "Cobblestone model" is "model kocich łbów"

REFERENCES

[1] Dietrych J., S. Kocańda, W. Korewa. 2006. *Podstawy Konstrukcji Maszyn (Basics of constructions of machines)*. Warsaw: Wydawnictwa Naukowo-Techniczne.

[2] Coulomb C. A..1785. *Théorie des machines simples en ayant égard au frottement de leurs parties et à la roideur des cordages (Theory of simple machines with focus on its parts' friction and line's stiffness).* Mém. des Mith. Phys. 161 – 342.

[3] Jaworski B., A. Dietłaf, L. Miłkowska, G. Siergiejew. 1984. *Kurs fizyki*, *Tom 1*, *Mechanika, podstawy fizyki cząsteczkowej i termodynamiki (Physics Course, Vol.1, Mechanics, foundations of particle physics and thermodynamics)*. Warsaw: PWN.

[4] Januszajtis A., J. Langer . 1987. *Fizyka – ilustrowana encyklopedia dla wszystkich (Physics – illustrated encyclopedia for everyone)*. Warsaw: WNT.

[5] Cross R. 2016. *Coulomb's Law for rolling friction*. Am. J. Phys. 84(3): 221–230.

[6] Wen *S., P. Huang P.* 2012. *Principles of Tribology*. John Wiley & Sons.

[7] Dupuit A. J. E. J. 1837. *Essai et expérience sur le tirage des voitures et sur le frottement de seconde espéce (The summary of work on the topic of vehicles' track and second order friction)* C. R. Aca. Sci. 9: 659 – 700,779.

[8] Kragelskii I. V., M. N. Dobychin, V. S. Kombalov. 1982. *Friction and wear – calculation methods*. Pergamon Press.

[9] Zapłata J., M. Pajor, G. Stateczny G. 2015. *Bezprzewodowy system kompensacji odkształceń cieplnych śrub pociągowych (Wireless system of compensation of heat deformations of lead screws).* Przegląd Mechaniczny 12: 38 – 41.

[10] PN-EN ISO 4287:1999: *Specyfikacje geometrii wyrobów - Struktura geometryczna powierzchni: metoda profilowa – Terminy, definicje i parametry struktury geometrycznej powierzchni (Specification of products' geometry – Geometric structure of the surface: profile method – terms, definitions and parameters of surface's geometric structure).*

Grzegorz M. Koczan
Warsaw University of Life Sciences
e-mail: grzegorz_koczan@sggw.edu.pl
(e-mail: gkoczan@fuw.edu.pl)

Juliusz Ziomek
Student at the University of Southampton
Department of Physical Sciences and Engineering
School of Electronics and Computer Science
Laureate of LXVII Polish National Physics Olympiad
e-mail: jkz1g18@soton.ac.uk